\documentclass[preprintnumbers, floatfix, letterpaper, twocolumn,aps,prd,epsfig,nofootinbib,natbib,longbibliography]{revtex4-1}
\usepackage{graphicx}
\usepackage{epstopdf}
\usepackage{latexsym}
\usepackage{amssymb}
\usepackage{amsmath}
\usepackage{color}
\usepackage{mathrsfs}
\usepackage{xparse}
\usepackage{float}
\usepackage[center]{subfigure}

\begin{document}

%%%%%%%%%%%%%%%%%%%%%%%%%%%%%%%%%%%%%%%%%%%%%%%%%%%%%%%%%%%%%%%
  \renewcommand\arraystretch{2}
 \newcommand{\bq}{\begin{equation}}
 \newcommand{\eq}{\end{equation}}
 \newcommand{\bqn}{\begin{eqnarray}}
 \newcommand{\eqn}{\end{eqnarray}}
 \newcommand{\nb}{\nonumber}
 \newcommand{\lb}{\label}
 \newcommand{\cb}{\color{blue}}
    \newcommand{\cc}{\color{cyan}}
        \newcommand{\cm}{\color{magenta}}
\newcommand{\rc}{\rho^{\scriptscriptstyle{\mathrm{I}}}_c}
\newcommand{\rd}{\rho^{\scriptscriptstyle{\mathrm{II}}}_c} 
\NewDocumentCommand{\evalat}{sO{\big}mm}{%
  \IfBooleanTF{#1}
   {\mleft. #3 \mright|_{#4}}
   {#3#2|_{#4}}%
}
\newcommand{\PRL}{Phys. Rev. Lett.}
\newcommand{\PL}{Phys. Lett.}
\newcommand{\PR}{Phys. Rev.}
\newcommand{\CQG}{Class. Quantum Grav.}
 %%%%%%%%%%%%%%%%%%%%%%%%%%%%%%%%%%%%%%%%%%%%%%%%%%%%%%%%%%%%%%%

\title{Langer Modification, Quantization condition and Barrier Penetration in Quantum Mechanics}

\author{Bao-Fei Li $^{1,2,3}$}
\email{libaofei@gmail.com}
\author{Tao Zhu $^{1,3}$}
\email{zhut05@zjut.edu.cn; Corresponding author}
\author{Anzhong Wang$^{4}$}
\email{Anzhong$\_$Wang@baylor.edu}
\affiliation{$^{1}$ \quad Institute for Theoretical Physics and Cosmology, Zhejiang University of Technology, Hangzhou, 310032, China\\
$^{2}$ \quad Department of Physics and Astronomy, Louisiana State University, Baton Rouge, Louisiana, USA\\
$^{3}$ \quad United Center of Gravitational Wave Physics (UCGWP), Zhejiang University of Technology, Hangzhou, 310032, China \\
$^{4}$ \quad GCAP-CASPER, Physics Department, Baylor University, Waco, TX 76798-7316, USA}

\date{\today}

\begin{abstract}

 The WKB approximation plays an essential role in the development of quantum mechanics and various important results have been obtained from it.  
 In this paper, we introduce another method, {\it the so-called uniform asymptotic approximations}, which is an analytical approximation method to  calculate the wave 
functions of the Schr\"odinger-like equations, and is applicable to various  problems, including cases with poles (singularities) and multiple turning points. An 
distinguished feature of the method is that in each order of the approximations  the upper bounds of the errors are given explicitly. By properly choosing the 
freedom introduced in the method, the errors can be minimized, which significantly improves the accuracy of the calculations. A byproduct of the method is to 
provide a very clear explanation of the Langer modification encountered in the studies of the hydrogen atom and harmonic oscillator. To further test our method, 
we calculate (analytically) the wave functions for several exactly solvable potentials of the Schr\"odinger equation, and then obtain the transmission coefficients 
of particles over potential barriers, as well as the quantization conditions for bound states.  We find that such obtained results agree with the exact ones extremely 
well.  Possible applications of the  method to other fields are also discussed.  

\end{abstract}

\maketitle

\section{Introduction}
\label{Intro}

%{\em Introduction}.---
A fundamental interest in quantum mechanics (QM) is to derive various physical quantities from the wave function of the  Schr\"{o}dinger equation. Due to the complexity
of the equation,  {it is extremely difficult to conduct   analytical analyses, and  various approximation methods  have been proposed.} Among them, the WKB approximation has played an essential role in the development of QM
and been widely used in many fields of physics and chemistry \cite{berry_semiclassical_1972,friedrich_working_2004, price_semiclassical_2018, hyouguchi_divergencefree_2002, Karnakov_wkb, froman_jwkb}.
In general,  { the one-dimensional Schr\"{o}dinger equation reads,}
\bqn
\lb{sch}
\frac{d^2\Psi(x)}{dx^2} + \frac{2m}{\hbar^2} \Big(E- V(x)\Big) \Psi(x)=0,
\eqn
which describes a particle of mass $m$ moving with total energy $E$ in a potential $V(x)$.  {Then, using the WKB approximations,   the  wave function $\Psi(x)$ can be approximately written in the form}
\bqn
\lb{WKB_function}
\Psi(x) \simeq \frac{\hbar}{\sqrt{2|p(x)|}} \exp{\left[\pm \frac{i}{\hbar} \int^x p(x') dx'\right]},
\eqn
where $p(x) = \sqrt{2m(E-V)}$ is the local momentum. The validity of the approximation  is restricted to the regions where the WKB condition is fulfilled,  
\bqn
\lb{WKB_condition}
\mathcal{Q}\equiv \hbar^2 \left|\frac{3 p'^2}{4 p^4} - \frac{p''}{2 p^3}\right| \ll 1.
\eqn
In this way, one can treat the reduced Planck constant $\hbar$ as a small parameter and extend the leading-order solution (\ref{WKB_function}) to high-orders.  

However, it is well known that the above condition can be violated or not fulfilled completely in many cases \cite{berry_semiclassical_1972}. For example, the WKB condition is always violated around  turning points
($p(x) = 0$), at which both $\mathcal{Q}$ and the WKB wave function (\ref{WKB_function})   diverge.   
In addition, the WKB condition is also violated around singular points (poles) of $p^2(x)$. For instance, for the radial Schr\"{o}dinger equation, the effective potential $V(r)$ contains a centrifugal term
\bq
\lb{potential}
V_C(r)=\frac{\hbar^2 l(l+1)}{2 m r^2},
\eq
which has   a second-order pole at the origin $r = 0$. Other typical potentials encountered in QM are presented in Table 1.

\begin{table}
\caption{\label{tab:table1}%
Some exactly solvable potentials and the choices of $q(x)$. The exact energy eignvalue spectra for these solvable potentials can be found in \cite{dong_wave_2011}.
}
\centering
\begin{tabular}{c|cc}
\toprule
\textrm{Potentials} & $V(x)$ & $q(x)$ \\
\hline
Hydrogen & $-\frac{e^2}{x} + \frac{\hbar^2 l(l+1)}{2 m x^2}$ &  $- \frac{1}{4 x^2} $\\
\hline
Harmonic oscillator & $\frac{1}{2} m \omega^2 x^2 + \frac{\hbar^2 l(l+1)}{2 m x^2}$ &$ - \frac{1}{4 x^2} $\\
\hline
Morse potential  & $v_0 e^{-2 \alpha x} + v_1 e^{-\alpha x} $ & $0$  \\
\hline
P\"{o}schl-Teller potential & $ \frac{v_0}{\cosh^2(\alpha x)}$ & $\frac{\alpha^2}{4 \cosh^2(\alpha x)}$  \\
\hline
Eckart potential & $ \frac{v_0}{ {\rm sinh}^2(\alpha x) }+ \frac{v_1}{ {\rm tanh}(\alpha x)}$ &  $ - \frac{\alpha^2}{4 {\rm sinh}^2(\alpha x)}  $\\
\toprule
%\bottomrule
\end{tabular}
\end{table}

Note that it is exactly because   of this second-order pole that the WKB approximation fails to give correct results for hydrogen atoms and harmonic oscillators \cite{young_wentzelbrillouinkramers_1930}. This  problem was studied by Langer
several decades ago, and shown that it   can be cured if one replaces $l(l+1)$ by $(l+1/2)^2$ in $V_C(r)$   \cite{langer_connection_1937}. This  modification now is considered  as a standard ingredient of the WKB method in QM \cite{berry_semiclassical_1972}. However, a rigorous and logically consistent derivation of this modification is still lacking.

Another situation that could violate the WKB condition is around the extreme point of $p^2(x)$ if $\mathcal{Q} \simeq |p''/2 p^3| \sim \mathcal{O}(1)$. This can arise in the  bound states with a potential well or in particle scattering with a potential barrier. In both cases the results from the WKB approximation becomes invalid if $\mathcal{Q}$ is not small enough at the extreme points.

 {To overcome these problems, various (approximate) methods have been proposed  \cite{dong_wave_2011}, including the complex  \cite{Zwaan29, Langer34, LL97, Berry90, DDP97} and uniform WKB methods \cite{langer_connection_1937, MG53,Dingle56, Alvarez04, DU14}. In the complex method, one first generalizes the above one-dimensional  problem into a complex $x$ plane, and then joins the two parts $x > a$ and $x < a$ by a path in the 
complex plane, which is sufficiently far from the turning point $x = a$, so the WKB solutions are valid along the path. With the seminal work of Delabaere, Dillinger and Pham \cite{DDP97}, the long-standing problem of the method, the ambiguities of the unique matching between the two parts  \cite{berry_semiclassical_1972}, was finally resolved. As a result, a simple and rigorous justification of the Zinn-Justin quantization condition and
its solution in terms of the ``multi-instanton expansion" for the symmetric double well problem \cite{ZJ02}, were derived. On the other hand, the uniform (in the sense of being smooth across the turning points)
 WKB approximation method is a generalization of the basic WKB approximations by employing the technique of comparison equations   \cite{Slavyanov96, FF96}. In this approach,   the  solution is mapped on to the one
  of a simpler equation which, however, has the same disposition of turning points as the original. Without bypassing  the turning points, as done in
 the complex method \cite{DDP97},    the uniform WKB approximations  also provide the wave function in the neighborhood of the turning points. Using this method, \'Alvarez was also able to derive the ``multi-instanton expansion" for the eigenvalues of the symmetric double well by using a Langer-Cherry uniform asymptotic expansion \cite{Alvarez04}. An excellent pedagogical introduction of this method was recently provided by Dunne and \"Unsal with both the double well and the Sine-Gordon potentials as concrete examples \cite{DU14}, in which 
the relations among the uniform WKB approach, multi-instanton and resurgence theory were also explored.}

 The purpose of this paper is  {to present another  approximation method, the so-called {\em uniform asymptotic approximation method}, developed systematically  by Olver \cite{olver1975, olver1997_book}. 
 Among other things, an
 essential  difference of this method from the above ones is that explicit bounds are constructed for the error terms associated with the approximations, although,
   similar to the uniform WKB method, it is also based on the comparison equations  \cite{Slavyanov96, FF96}.
 In particular, we shall calculate accurately  the wave function of the Schr\"{o}dinger equation (\ref{sch}) with singular potentials such as those given in Table 1, although the method is quite general, and in principle 
 can be applied to any problem of
 the Schr\"{o}dinger equation (\ref{sch}), including the cases with poles and multiple turning points. This method 
 has been already }  {shown to be powerful and robust when applied to calculations} of the mode functions and primordial power spectra in a variety of slow-roll inflationary models  \cite{Habib:2002yi, Habib:2004kc, Wang10, Zhu:2014wfa, AKN16, Wu:2017joj},  { adiabatic regularization of the power spectra in various inflationary models \cite{Alinea:2015pza, Alinea:2016qlf}, }inflation with nonlinear dispersion relations \cite{Zhu:2013fha, Zhu:2013upa, Zhu:2014wda, Zhu:2014aea, Zhu:2016srz, Qiao:2018dpp, Ding:2019nwu, Qiao:2019hkz},  and loop quantum cosmology \cite{Zhu:2015ata, Zhu:2015owa, Zhu:2015xsa, Li:2018vzr}, as well as studying the parametric resonance during inflation and reheating \cite{Zhu:2018smk}.  {As mentioned above, } the major advantage of the method is that the errors in each order of approximations can be estimated and the upper bounds of the errors are always known.  In particular, for certain models, it was found that the errors are no larger than $0.15\%$ up to the third-order of approximations  \cite{Zhu:2016srz}.

 In the application of this method to hydrogen atoms and harmonic oscillators, we provide a rigorous derivation of the Langer modification and expressions for the quantization conditions and the barrier transmission coefficients.  Applications of our method to some well-known examples  are also presented, in order to further test it.

 \section{Uniform asymptotic approximation method}
%{\em Uniform asymptotic approximation method}.---

\subsection{Wave functions in the uniform asymptotic approximation}

Let us first write the standard form of (\ref{sch}) in the form \cite{olver1997_book, olver1975},
\bqn\lb{eom}
\frac{d^2\Psi(x)}{dx^2} = \left\{g(x)+q(x)\right\} \Psi(x),
\eqn
where $g(x)+q(x)=- p^2(x)/\hbar^2$. Note that, for any given $p(x)$, here we introduce two functions $g(x)$ and $q(x)$, and to fix them uniquely,  we require that the errors in each order of approximations be minimized. This is one of the major  {ingredients} of the method. Then, such defined $g(x)$ in general can have zero points $g(x_i) = 0$, which are called {\em turning points} in the uniform asymptotic approximation. Except such points, $g(x)$ may also have other types of transition points, such as poles and extreme points. According to the theory of the second-order ordinary differential equations \cite{olver1997_book, olver1975}, the wave function $\Psi(x)$ sensitively depends on the number and nature of turning points, poles and extreme points. Analyzing the corresponding error control function  around poles and extreme points  provides the main guidance on how to determine   $g(x)$ and $q(x)$ \cite{olver1997_book, olver1975}. In addition, around each of the turning points $x_i$, we require $|q(x)| \ll |g(x)/(x-x_i)|$, while away from them we require  $|q(x)| \ll |g(x)|$ \cite{olver1997_book, olver1975}.

At the turning points, the WKB condition (\ref{WKB_condition}) is violated, and the wave function (\ref{WKB_function}) becomes invalid. Generally the turning points can have different types, depending on the nature of the zeros of $g(y)$. The simplest case is the single turning point which denotes the simple and single zero of $g(y)$. Beside this, $g(y)$ may also have a pair of turning points, which could be both real and single, double, or even complex conjugated. In the following we are going to construct the uniform wave function for above two cases individually.

\subsubsection{For single turning point}

Let us first consider the single turning point, say, denoted by $x_0$. We can write the function $g(x)$ in the form of
\bqn
g(x) = f(x)(x-x_0),
\eqn
where $f(x)$ is a regular function. In quantum mechanics, this turning point can be thought as the critical point that separates the  {classically} allowed and forbidden regions of a quantum system if $q(x)=0$. When we have to chose $q(x) \neq 0$, it will produce a small shift on this point. Around the turning point $x_0$,
  the   wave function $\Psi(x)$ can be written as \cite{olver1997_book}
\bqn\lb{wave1}
\Psi(x)= \left(\frac{\xi(x)}{g(x)}\right)^{1/4} \Bigg(a_0 {\rm Ai}(\xi) + b_0 {\rm Bi}(\xi)\Bigg),
\eqn
where ${\rm Ai}(\xi)$ and ${\rm Bi(\xi)}$ are the Airy functions, $a_0$ and $b_0$ are two integration constants, and $\xi(x)$ is chosen to be a monotonic function of $x$, which has the same sign of $g(x)$ with relations $\sqrt{|\xi|}d\xi= \sqrt{|g(x)|}dx$ and $\xi(x_0)=0$. The errors  are controlled by the error control function,  
\bqn
&&\mathscr{H}(\xi) \equiv \frac{5}{24 |\xi|^{3/2}} - \int_{x_0}^{x} \Bigg\{\frac{q(\tilde x)}{g(\tilde x)} - \frac{5 g'^2(\tilde x)}{16 g^3(\tilde x)} + \frac{g''(\tilde x)}{4g^2(\tilde x)}\Bigg\} \nb\\
&&  ~~~~~~~~~~~~~~~~~~~~~~~~~~~~~~ \times \sqrt{|g(\tilde x)|} d\tilde x.
\eqn

\subsubsection{A pair of turning points}

For a pair of turning points $x_1$ and $x_2$, they could be: (1) both real and different $x_1 \neq x_2$; (2) both real but equal  $x_1=x_2$; or (3) complex conjugate $x_1=x_2^*$.  In each case, between these points, $g(y)$ usually has one extreme.

If this extreme is a minimal point of $g(y)$, we can construct the wave function $\Psi(x)$ in terms of the parabolic cylinder functions $U(-\zeta_0^2/2, \sqrt{2}\zeta)$ and $\bar U(-\zeta_0^2/2 \sqrt{2}\zeta)$ as \cite{olver1975},
\bqn\lb{wave2}
\Psi(x) &=&\left(\frac{\zeta^2-\zeta_0^2}{- g(x)}\right)^{1/4} \Big(a_1 U(-\zeta_0^2/2, \sqrt{2}\zeta)  \nb\\
&& + b_1 \bar U(-\zeta_0^2/2, \sqrt{2}\zeta)\Big).
\eqn
If the extreme is a maximal  point, the wave function can be also constructed in terms of  parabolic cylinder functions,  but now in terms of $W(\zeta_0^2/2, \sqrt{2}\zeta)$ and $W(\zeta_0^2/2, -\sqrt{2}\zeta)$, 
\bqn\lb{wave3}
\Psi(x) &=&\left(\frac{\zeta^2-\zeta_0^2}{- g(x)}\right)^{1/4} \Big(a_2 W(\zeta_0^2/2, \sqrt{2}\zeta) \nb\\
&& + b_2 W(\zeta_0^2/2, -\sqrt{2}\zeta)\Big).
\eqn
In both cases, the variable $\zeta(x)$ is a monotonic function of $x$ and  defined by   $\sqrt{|\zeta^2-\zeta_0^2|}d\zeta=\sqrt{|g(x)|}dx$ with $\zeta(x_1)=- |\zeta_0|$, $\zeta(x_2)=|\zeta_0|$,  and $\zeta_0^2 = \pm (2/\pi) |\int_{x_1}^{x_2} \sqrt{|g(x)|}dx|$, where   $``+"$ $(``-")$ corresponds to the case in which  the two turning points $x_1$ and $x_2$ are  real (complex conjugate). The associated error control function of the above two wave functions is,  
\bqn\lb{error_I}
\mathscr{I}(\zeta)&=& \int_{\pm \zeta_0}^\zeta \left[\frac{5 \zeta_0^2}{4 |v^2-\zeta_0^2|^{5/2}} - \frac{3 }{4 |v^2-\zeta_0^2|^{3/2}}\right]dv\nb\\
&& - \int_{x_{1, 2}}^{x} \Bigg\{\frac{q(\tilde x)}{g(\tilde x)} - \frac{5 g'^2(\tilde x)}{16 g^3(\tilde x)} + \frac{g''(\tilde x)}{4g^2(\tilde x)}\Bigg\} \sqrt{|g(\tilde x)|} d\tilde x.\nb\\
\eqn

The wave functions given in (\ref{wave1}), (\ref{wave2}) and (\ref{wave3}) are   valid in the neighborhoods of the turning points $x_i$. The extension of them beyond these points crucially depends  on the behaviors of the corresponding error control functions in the extended regions. In the following let us consider it   for the case with a second-order pole, as shown in (\ref{potential}).  

\subsection{Second-order pole and Langer modification}

As mentioned  {above},  for the radial Schr\"{o}dinger equation, the effective potential $V_C(r)$ given by (\ref{potential})  contains a second-order  pole  at the origin, at which
we have $(\xi, \; \zeta) \to \pm \infty$, and
\bqn
(\mathscr{H},\; \mathscr{I}) \to - \int^x \left(\frac{q}{g}- \frac{5 g'^2}{16 g^3} + \frac{g''}{4 g^2}\right)\sqrt{|g|}dx.
\eqn
 {Since near the second-order pole,  $g(x)$ has the asymptotic behavior}
\bq
\lb{asymp}
g(x) \sim a/x^2,
\eq
we find that  the error control function  takes the limit  
\bqn
\lb{error}
\left(\mathscr{H},\; \mathscr{I}\right) \to  \lim_{x \to 0} \left(- \frac{\ln x}{4 \sqrt{|a|}} - \int^x \frac{q}{\sqrt{|g|}}dx \right).
\eqn
 { In order to make our analytic solutions in (\ref{wave1}), (\ref{wave2}) and (\ref{wave3})   also valid at the origin, we must  require  that the error control function given by  (\ref{error}) are finite at the pole $x=0$. Using the asymptotic behavior of $g(x)$ given in (\ref{asymp}), it is straightforward to show that  the right-hand side of (\ref{error})  can be made finite at the pole as long as \cite{Zhu:2013upa}}
\bq
\lb{qx}
q(x) = - \frac{1}{4 x^2},
\eq
which is nothing but exactly  the Langer modification.
 Thus,  the latter  is simply  the result of imposing the condition that  the error control function be finite at the pole in the uniform asymptotic approximation method.

\subsection{Extreme point and the elimination of the error term}

The extreme points of $g(x)$ in general arise from quantum system with a potential well or barrier in the region between two turning points. These extreme points are  the same as the bottom of the well or the top of the potential if one chooses $q(x)=0$. As mentioned above, the existence of the extreme points will make the WKB approximation invalid if $\mathcal{Q}$ is not small enough at the extreme. Such cases   rise when the potential wells or barriers are sharply peaked, for which $g(x)$ has two coalescent turning points $x_1$ and $x_2$. To  {be more specific},  
let us write $g(x)$ in the form
\bqn
g(x) = f(x)(x-x_1)(x-x_2),
\eqn
where $f(x)$ is a finite and regular function with $f(x_i) \not= 0$. Then, we expect the dominant contribution to the integral of (\ref{error_I}) arises from the lower limit.  Therefore, we can formally expand the error control function $\mathscr{I}(\zeta)$ about the turning points and find \footnote{ {It should be noted that the integration of the form, $I = \int{dx g(x) \exp\left[i f(x)\right]}$, by using the method of stationary phase, is well-established, see, for example, \cite{Miller70}, where there are two roots, $x_1$ and $x_2$,  of the equation $f'(x) = 0$. When   $x_2-x_1$ is small, the method leads to the solution,  $I = g_0 \exp(i f_0)2\pi \left|2/f'''_0\right|^{1/3}Ai[f_0(2/f'''_0)^{1/3}]$, where
$x_0$ is the value of $x$ for which $f''(x_0) = 0$,  $g_0 \equiv g(x_0)$ and so on. However, to our current purpose, we find that the expression of (\ref{errorA}) is more suitable.}},  
\bqn\lb{errorA}
\mathscr{I}(\zeta) &\simeq& \left.\frac{7 f'^2- 6 f f''}{32 |f|^{5/2}}\right|_{{\rm Re}(x_{i})} \ln|x_2-x_1|  \nb\\
&& - \int_{{\rm Re}(x_{i})} \frac{q}{\sqrt{|g|}}dx, \; (i = 1, 2),
\eqn 
 {where ${\rm Re}(x_{i})$ denotes the real part of the turning point $x_i$}.
Note that in deriving the above expression we had ignored the small corrections. We also note  that when the turning points $x_1$ and $x_2$ are  {close}  to each other, the dominant contributions to the error control function come from the $\ln|x_2-x_1|$ term, which becomes divergent at  $x_1=x_2$. It is somehow surprising that   such dominant contributions seemingly had never  been noticed before,
 and later they will play an essential role in determining the extension of the wave functions to the regions near the extreme point.

With the knowledge of the error control function (\ref{errorA}), we are now in a position to  eliminate the dominant error term in (\ref{errorA}) by properly choosing $q(x)$ in the second term of (\ref{errorA}). To achieve this, we expand $q(x)$ at one of the turning points \footnote{ {It should be noted that in general the expansion should be carried out in terms of $x_i$  in the complex $x$ plane \cite{LL97}. However, we find that for the analysis of the error control function $\mathscr{I}(\zeta)$ defined by (\ref{errorA}) near the  turning point, the expansion  alone the real axis is sufficient. In particular, it is involved only with the choice of the zeroth-order term $q_0$, as can be seen from (\ref{q0}) and (\ref{q0V}).}},
\bqn
q(x) \simeq q_0 + q_1 (x-{\rm Re}(x_{i})).
\eqn
Then the elimination of the first error term in (\ref{errorA}) requires
\bqn\lb{q0}
q_0 =  \left.\frac{7 f'^2- 6f f''}{32  f^2}\right|_{{\rm Re}(x_{i})}.
\eqn
This  represents one of the important conditions for the choice of the function $q(x)$. Additional requirements include that $|q(x)|$  be negligible  { in comparison with} $|g(x)|$ in the regions that is away from the extreme and turning points. Then, we expect that the right hand side of (\ref{q0}) does not contain $q_0$ and should be independent of the nature of the two turning points $x_1$ and $x_2$. For this requirement, one can relate $q_0$ to the derivatives of the function $g(x)$ at the extreme point $x_m$ via the relation,
\bqn\lb{q0V}
q_0 =  \left(\frac{7g'''^2}{288 g''^2}- \left.\frac{g''''}{32 g'' }\right)\right|_{x_m}.
\eqn

\section{Improved quantization conditions and potential barrier transmission coefficients}

With the above considerations, now we are at the position to generalize the wave functions near poles and extreme points. These wave functions then can be utilized to derive the quantization conditions for bound states or the quantum  {transmission coefficients} of a particle  through a potential barrier.  In this section, we are going to discuss quantization conditions and potential barrier transmission individually by using the quantum mechanical wave functions derived in the above section with the corresponding $q(x)$ calculated for specific potentials.

\subsection{Improved Quantization Condition}

One of  {the most important topics} in quantum mechanics is to determine the bound states and energy eigenvalue spectra for a quantum mechanical system.   {As mentioned in the introduction, the conventional WKB method fails to describe} the correct behaviors of wave functions around the poles, extreme points, and turning points of $E-V(x)$ in the Schr\"{o}dinger equation (\ref{sch}). Especially at the turning point, the WKB wave function (\ref{WKB_function}) becomes divergent. One approach to avoid this singularity is the complex method in which the WKB approximation has been extended to the complex $x$ plane. In this way, the two WKB wave functions in the classically allowed and forbidden regions can be connected without the knowledge of the wave functions around the turning points. With this procedure, when a particle is trapped in a potential well $V(x)$, the corresponding energy eigenvalue spectra can be obtained by using the conventional WKB quantization condition,
\bqn\lb{BS}
\int_{\tilde x_1}^{\tilde x_2} \sqrt{\frac{2 m}{\hbar^2} \left(E - V(x)\right)}dx = \left(n+\frac{1}{2}\right) \pi,
\eqn
where $\tilde x_1$ and $\tilde x_2$ are two zeros of $E-V(x)$ (or classical turning points of (\ref{sch})) and $n$ denotes the quantum number. It is worth noting that $\tilde x_1$ and $\tilde x_2$ is slightly different from the turning points $x_1$ and $x_2$ of $g(y)$ with $q(x) \neq 0$. For one single turning point system, where a particle is trapped in a classically allowed region with a boundary at $x_b$, the WKB quantization gives,
\bqn
\int_{x_b}^{\tilde x_0} \sqrt{\frac{2 m}{\hbar^2} \left(E - V(x)\right)}dx = \left(n+\frac{3}{4}\right) \pi.
\eqn

 {However, in   the uniform asymptotic approximations,  } if $g(x)$ has two turning points $x_1$ and $x_2$,  the wave function (\ref{wave2}) leads to the quantization condition,
\bqn\lb{iBS}
\frac{\pi \zeta_0^2}{2}=\int_{x_1}^{x_2} \sqrt{-g(x)} = \left(n+\frac{1}{2}\right)\pi.
\eqn
When $g(x)$ has only one turning point $x_0$, the wave function (\ref{wave1}) leads to the quantization condition,
\bqn
\int_{x_0}^{x_b} \sqrt{-g(x)} = \left(n+\frac{3}{4}\right)\pi,
\eqn
where $x_b$ is the boundary of the classically allowed region. The above quantization condition is different from that derived in the WKB approximation except for the case with $q(x)=0$. For the quantum mechanical system that $q(x)$ has to be nonzero, the above quantization condition can provide a significant improvement on the calculations of the energy eigenvalue spectra for a quantum mechanical system. In the following, for applications of the above  quantization condition, we consider several representative exactly solvable systems, including hydrogen atoms, harmonic oscillators in D dimensions, Morse potential, P\"{o}schl-Teller potential, and Eckart potential, as given in Table.~\ref{tab:table1}.

\subsubsection{Hydrogen atoms}

For hydrogen atoms, the electron moves in the atom with a Coulomb potential and a centrifugal term \cite{dong_wave_2011},
\bqn
V(x) = - \frac{e^2}{x} + \frac{\hbar^2 l(l+1)}{2 m x^2},
\eqn
where $e$ denotes the electron charge. The Schr\"{o}dinger equation (\ref{sch}) of the hydrogen atom can be solved exactly, and from the exact solution one can easily obtain the exact energy eigenvalue spectra as \cite{dong_wave_2011}
\bqn
E_{n} = - \frac{m e^4}{2 \hbar^2 (n+l+1)^2}.
\eqn
If we apply the WKB quantization condition (\ref{BS}), the WKB energy eigenvalue spectra read
\bqn
E_n = - \frac{m e^4}{2 \hbar^2 (n+ 1/2 + \sqrt{l(l+1)})^2},
\eqn
which does not match the exact result. A standard procedure to solve this problem in quantum mechanics can be implemented using the Langer modifciation by simply replacing $l(l+1)$ by $(l+1/2)^2$ in the Schr\"{o}dinger equation \cite{langer_connection_1937}.

In the uniform asymptotic approximation, the above problem can be resolved by properly choosing $q(x)$ to eliminate  the divergent errors of the approximate solution arising from the existence of the second-order pole and 
the extreme point.  It can be seen that the Schr\"{o}dinger equation (\ref{sch}) for the hydrogen atom has a second order pole, one extreme point, and two turning points. Then, according to the analysis in the above section, $q(x)$ has to be chosen such that
\bqn
q(x) \to  - \frac{1}{4 x^2} \;\;\; {\rm as}\;\;\; x \to 0, \lb{AQ}\\
q(x_m) = \left(\frac{7g'''^2}{288 g''^2}- \left.\frac{g''''}{32 g'' }\right)\right|_{x_m}. \lb{BQ}
\eqn
To meet these two conditions, $q(x)$ must be
\bqn \lb{hy_q}
q(x) = - \frac{1}{4 x^2}.
\eqn
With this choice, the two turning points $x_1$ and $x_2$ can be solved from $g(x)=0$, yielding
\bqn
x_1 = - \frac{e^2}{2 E} + \frac{\sqrt{m^2 E^4 + m E (l+1/2)^2 \hbar^2}}{2 m E}, \\
x_2 = - \frac{e^2}{2 E} - \frac{\sqrt{m^2 E^4 + m E (l+1/2)^2 \hbar^2}}{2 m E}.
\eqn
Substituting them into the improved quantization condition (\ref{iBS}), one immediately obtains
\bqn
E_n = - \frac{m e^4}{2 \hbar^2 (n+l+1)^2},
\eqn
which precisely recovers the exact result. The eigenstate for each quantum number then is given by the wave function in (\ref{wave2}) with $q(x)$ given in (\ref{hy_q}), which is uniformly valid in the whole region $x \in (0, +\infty)$. It is worth noting that the above analysis can be easily extended to the hydrogen-like atoms.

\subsubsection{Harmonic Oscillators}

 {The potential for the  harmonic oscillator in D dimensions is given by \cite{dong_wave_2011}
\bqn
V(x) &=& \frac{1}{ 2} m^2 \omega^2 x^2 + \frac{\hbar^2 }{2 m x^2}\Bigg(l \left(D+l-2\right)\nb\\
&& +\frac{(D-1)(D-3)}{4}\Bigg).
\eqn
}
The exact solution of the Schr\"{o}dinger equation (\ref{sch}) with the above potential leads to  \cite{dong_wave_2011} {
\bqn
E_n = \left(2n +  l + \frac{D}{2} \right) \hbar \omega.
\eqn}
If we perform the WKB quantization condition (\ref{BS}), the WKB energy eigenvalue spectra read {
\bqn
E_n &=&  \Bigg(2n +  \sqrt{l \left(D+l-2\right)+\frac{(D-1)(D-3)}{4}} \nb\\
&& ~~~~~ + 1 \Bigg) \hbar \omega.
\eqn}
Similar to the hydrogen atom, it does not match the exact result.

To do the analysis in the uniform asymptotic approximation, we observe that Schr\"{o}dinger equation (\ref{sch}) for the harmonic oscillator  {in D dimensions} has one second-order pole, one extreme point, and two turning points. This is very similar to the case in the hydrogen atom. Then the choice of $q(x)$ has to meet the two conditions in (\ref{AQ}) and (\ref{BQ}), which leads to
\bqn
q(x) = - \frac{1}{4 x^2}.
\eqn
With this choice, the improved quantization condition (\ref{iBS}) then gives {
\bqn
E_n =  \left(2n +  l + \frac{D}{2} \right) \hbar \omega,
\eqn}
which again recovers the exact result. Similarly the eigenstate for each quantum number is given by the wave function in (\ref{wave2}) which is valid in the whole region of interest.

\subsubsection{Morse potential}

The Morse potential is given by \cite{dong_wave_2011}
\bqn
V(x) = v_0 e^{-2 \alpha x} + v_1 e^{-\alpha x},
\eqn
where $v_0$, $v_1$, and $\alpha$ are three parameters in the potential. The exact solution of the Schr\"{o}dinger equation (\ref{sch}) with the Morse potential leads to the following energy eigenvalue spectra \cite{dong_wave_2011},
\bqn
E_n &=&  - \frac{1}{8 m v_0}\Bigg[2 m v_1^2 + (2n+1)\alpha \hbar \Big(2 \sqrt{2 m v_0} v_1 \nb\\
&& +(2n+1) v_0 \alpha \hbar\Big)\Bigg].
\eqn
If we perform the WKB quantization condition (\ref{BS}),  the WKB energy eigenvalue spectra takes exactly the same form as the exact result. This  is in contrast to the cases for the hydrogen atom and   {the harmonic oscillator in D dimensions}.

The analysis in the uniform asymptotic approximation can give the same result as well. Observing that the Morse potential has one extreme point between two turning points, application of  the criterion (\ref{q0V}) immediately leads to
\bqn
q(x_m) =  \left(\frac{7g'''^2}{288 g''^2}- \left.\frac{g''''}{32 g'' }\right)\right|_{x_m} =0.
\eqn
For this case once we choose $q(x)=0$, and then  the improved quantization condition (\ref{iBS}) recovers the WKB quantization condition (\ref{BS}).

\subsubsection{P\"{o}schl-Teller potential}

The P\"{o}schl-Teller potential is given by \cite{dong_wave_2011}
\bqn
V(x) = \frac{v_0}{\cosh^2(\alpha x)},
\eqn
where $v_0 <0$ and $\alpha$ are two parameters describing the potential. The exact solution of the Schr\"{o}dinger equation (\ref{sch}) with the P\"{o}schl-Teller potential leads to the exact energy eigenvalue spectra \cite{dong_wave_2011},
\bqn
E_n &=& v_0 - \frac{\alpha^2 \hbar^2}{4 m} \Bigg[2 n^2+2 n+1 \nb\\
&& - (2n+1)\sqrt{1-\frac{8 m v_0}{\alpha^2 \hbar^2}}\Bigg].
\eqn
If we perform the WKB quantization condition (\ref{BS}), the WKB energy eigenvalue spectra read
\bqn
E_n &=&  v_0 - \frac{\alpha^2 \hbar^2}{4 m} \Bigg[\frac{(2 n+1)^2}{2}\nb\\
&& - (2n+1)\sqrt{-\frac{8 m v_0}{\alpha^2 \hbar^2}}\Bigg].
\eqn
Obviously it does not match the exact result.

Considering that the P\"{o}schl-Teller potential only contains one extreme point between the two turning points, we can use the the criterion (\ref{q0V})  to determine the choice of $q(x)$. An immediately calculation shows that
\bqn
q_0 =  \frac{\alpha^2}{4}.
\eqn
As a result, a proper choice for $q(x)$ is
\bqn
q(x) = \frac{\alpha^2}{4 \cosh^2(\alpha x)}.
\eqn
With this choice, the improved quantization condition (\ref{iBS}) gives
\bqn
E_n &=&  v_0 - \frac{\alpha^2 \hbar^2}{4 m} \Bigg[2 n^2+2 n+1 \nb\\
&& - (2n+1)\sqrt{1-\frac{8 m v_0}{\alpha^2 \hbar^2}}\Bigg],
\eqn
which again is precisely the same as the exact result. Similarly the eigenstate for each quantum number is given by the wave function in (\ref{wave2}) which is valid in the whole region of interest.

\subsubsection{Eckart potential}

The Eckart potential takes the form \cite{dong_wave_2011}
\bqn
V(x) = \frac{v_0}{\sinh^2(\alpha x)} + \frac{v_1}{\tanh(\alpha x)},
\eqn
where $v_0$, $v_1$, and $\alpha$ are three parameters describing the potential. The exact solution of the Schr\"{o}dinger equation (\ref{sch}) with the Eckart potential leads to the exact energy eigenvalue spectra, which reads \cite{dong_wave_2011}
\bqn
E_n &=&  - \frac{2 m v_1^2}{\alpha^2 \hbar^2 (\sqrt{1+8 m v_0/(\alpha^2 \hbar^2)} + 2n +1)^2} \nb\\
&& - \frac{\alpha^2 \hbar^2 (\sqrt{1+8 m v_0/(\alpha^2 \hbar^2)} + 2n +1)^2}{8 m}.
\eqn
If we perform the WKB quantization condition (\ref{BS}), the WKB energy eigenvalue spectra read
\bqn
E_n &=&  - \frac{2 m v_1^2}{\alpha^2 \hbar^2 (\sqrt{8 m v_0/(\alpha^2 \hbar^2)} + 2n +1)^2} \nb\\
&& - \frac{\alpha^2 \hbar^2 (\sqrt{8 m v_0/(\alpha^2 \hbar^2)} + 2n +1)^2}{8 m}.
\eqn
We observe that it does not match the exact result.

Since the Eckart potential contains one second-order pole at $x=0$ and one extreme point between the two turning points, we need to use the two criterions (\ref{qx}) and (\ref{q0V}) to determine the choice of $q(x)$.  {A straightforward calculation} shows that a proper choice for $q(x)$ to meet both conditions is
\bqn
q(x) = - \frac{\alpha^2}{4 \sinh^2(\alpha x)}.
\eqn
With this choice, the improved quantization condition (\ref{iBS}) gives
\bqn
E_n& =& - \frac{2 m v_1^2}{\alpha^2 \hbar^2 (\sqrt{1+8 m v_0/(\alpha^2 \hbar^2)} + 2n +1)^2} \nb\\
&& - \frac{\alpha^2 \hbar^2 (\sqrt{1+8 m v_0/(\alpha^2 \hbar^2)} + 2n +1)^2}{8 m},
\eqn
which again is precisely the same as the exact result. Similarly the eigenstate for each quantum number is given by the wave function in (\ref{wave2}), which is valid in the whole region of interest.

In summary, it can be seen that from the above five examples that,  {\em while the standard WKB quantization condition fails to predict correct energy eigenvalues $E_n$ except for the Morse potential, the quantization condition (\ref{iBS})  yields precisely the exact results for all these potentials}. We note that the exact energy eigenvalue spectra for these potentials can also be obtained from alternative approaches, for examples, by using the supersymmetry quantum mechanics \cite{cooper_supersymmetry_2001} or the proper quantization rules developed in \cite{qiang_proper_2010, dong_wave_2011, serrano_qiang_2010}. In contrast to these approaches for solvable potentials, we would like to mention here that the improved quantization condition (\ref{iBS}) is not only applicable to the solvable potentials, but also valid for more general potentials that do not admit exact solutions. More importantly, the bound states for a general potential can also be obtained from the wave functions (\ref{wave2}) which are derived from our approach.

\subsection{Potential Barrier Transmission}

 {In the above subsection, we have  studied } the bound state and energy eigenvalue problems when a particle moves in a potential well. In this subsection, let us move on to the scattering problem for a particle passing through a potential barrier. As mentioned in the above, the standard WKB approximation cannot give the correct behaviors of the wave function about the turning points. When the energy of the particle is below the peak of the barrier, the Schr\"{o}dinger equation (\ref{sch}) normally has two real turning points $\tilde x_1$ and $\tilde x_2$. For this case, the complex method of the WKB approximation leads to an approximate transmission coefficient,
\bqn
T \simeq \exp{\left(- 2 \int_{\tilde x_1}^{\tilde x_2} \sqrt{- \frac{2m}{\hbar^2} (E - V(x))} dx \right)},
\eqn
where $\tilde x_1$ and $\tilde x_2$ are two turning points defined by $E-V(x)=0$.

The transmission coefficient in the uniform asymptotic approximation can be directly obtained from the wave function given in (\ref{wave3}) with known $q(x)$. This treatment is universal and does not depend on whether the energy of the particle is above or below the peak of the barrier. Using the wave function (\ref{wave3}),  the improved transmission coefficient reads
\bqn\lb{transmission}
T = \frac{1}{1+e^{\pi \zeta_0^2}}= \left[1+\exp\left(2 \int_{x_1}^{x_2}\sqrt{g(x)}dx\right)\right]^{-1}. ~~~~
\eqn
We note that $\zeta_0^2$ is positive when $x_1$ and $x_2$ are real and negative when $x_1$ and $x_2$ are complex conjugated. It is worth emphasizing again that (\ref{transmission}) is valid  {in either case in which }the energy $E$ is above or below the peak of the potential barriers.

\begin{figure*}
%\subfigure[]{\label{gofy1}
\includegraphics[width=8.1cm]{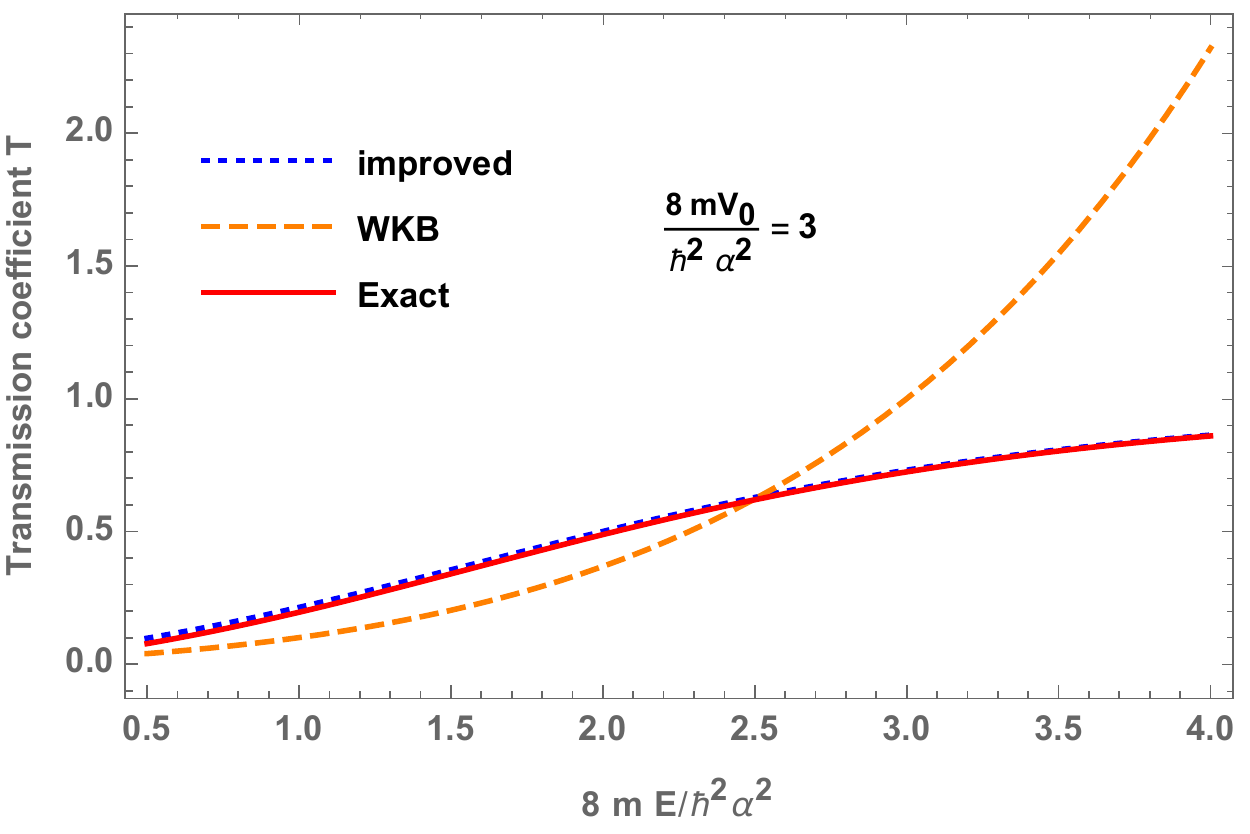}
%\subfigure[]{\label{gofy2}
\includegraphics[width=8.1cm]{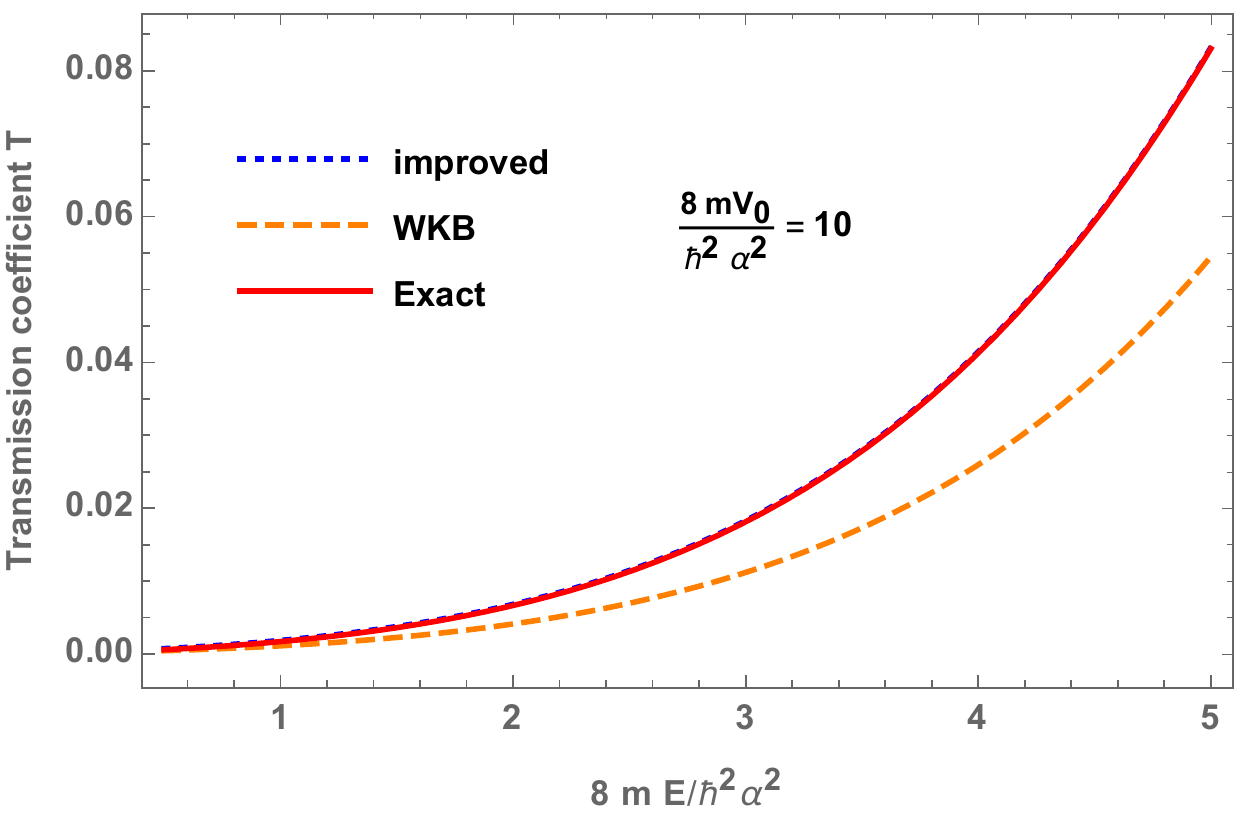}
\caption{Comparison among the exact transmission coefficient $T$, WKB transmission coefficient (\ref{WKB_trans}), and our ``improved" transmission coefficient (\ref{improved_trans}).} \label{fig:figure1}
\end{figure*}

To calculate the transmission coefficient (\ref{transmission}) for a specific form of the potential, we first need to determine the choice of $q(x)$. In this subsection, instead of calculating various  potentials in details,    let us consider the P\"{o}schl-Teller potential barrier with  a positive $v_0$ as an example. We note that this potential barrier has been applied to the calculation of the black hole quasi-normal modes \cite{berti_quasinormal_2009, konoplya_quasinormal_2011} and the recent studying of the primordial perturbation during quantum bounce in the loop quantum cosmology \cite{Wu:2018sbr, Wu:2018sbr1, Wu:2018sbr2, Wu:2018sbr3}. The choice of $q(x)$ is the same as in the case for the P\"{o}schl-Teller potential well. Then, the  transmission coefficient (\ref{transmission}) reads
\bqn\lb{improved_trans}
T = \frac{1}{1+e^{\pi (\sqrt{8 mv_0/\hbar^2 \alpha^2- 1} - \sqrt{8mE/\hbar^2\alpha^2 })}},
\eqn
while the (standard)  WKB approximation gives \cite{gottfried}
\bqn\lb{WKB_trans}
T = e^{- \pi (\sqrt{8 mv_0/\hbar^2 \alpha^2} - \sqrt{8mE/\hbar^2 \alpha^2})}.
\eqn
It is worth noting that the choice of $q(x)$ is only possible when $8mv_0/\hbar^2 > \alpha^2$ for positive $v_0$. When $8mv_0/\hbar^2 < \alpha^2$, as mentioned above, the choice of $q(x)$ changes the properties of turning points significantly, which makes the approximation not applicable. In Fig.~\ref{fig:figure1}, we present the  transmission coefficient (\ref{improved_trans}), the WKB transmission coefficient (\ref{WKB_trans}) and the exact result, from which one can see  that the   transmission coefficient (\ref{improved_trans}) fits the exact result extremely well.

\section{Summary and Outlook}

 {In this paper,  a new analytical approximation method to solve the Schr\"{o}dinger equation,  {the so-called {\em uniform asymptotic approximation method}, developed systematically  by Olver \cite{olver1975,olver1997_book}, }
 is presented and applied to several cases of interest in quantum mechanics. This new method makes use of the technique of uniform approximations to map the original Schr\"{o}dinger  equation to a simpler equation for which an approximate  solution can be found analytically. }  {  One of the major advantage of the method is that the errors in each order of approximations can be estimated and their upper bounds  are always known explicitly.  In particular, for certain models, it was found that the errors are no larger than $0.15\%$ up to the third-order of approximations  \cite{Zhu:2016srz}.}

 { To illustrate the above, in this paper we have given the general solutions from the new method in the region where one single  turning point or two turning points are encountered.  One of the advantages of the new method is that by carefully choosing the function $q(x)$, the errors of the approximate solutions can be well-controlled. In general,  the function $q(x)$ has to be chosen to satisfy two conditions:  
(a) near the turning  points $x_i$, $|q(x)|\ll |g(x)/(x-x_i)|$; and 
(b) away from the turning points,  $|q(x)|\ll|g(x)|$.  
Moreover, when one of the the boundaries of the interesting region, for example $x=0$, becomes a second-order pole, $q(x)$  has to be fixed correspondingly to  $-\frac{1}{4x^2}$ in order to make the error control function finite at the pole. One of the results from this particular choice of the function $q(x)$ is that the quantization conditions for bound states from the exact solutions can be recovered from our approximate solutions for several quantum mechanical systems. Moreover, when the form of the error control function can be analytically expressed in terms of $q(x)$ and $g(x)$, the function $q(x)$ can also be chosen to minimize  the errors at the tuning point or at the extreme point. With the new method, we have also derived  the transmission coefficients for particle scattering over a potential barrier, which  significantly improve the accuracy of the results obtained from the standard WKB approximations. }  

 {It must be noted that in this article we have mainly compared the uniform asymptotic approximation method  introduced in this paper with the conventional (standard) WKB approximations. Other methods, such as the  complex \cite{Zwaan29,Langer34,LL97,Berry90,DDP97} and uniform WKB  \cite{langer_connection_1937,MG53,Dingle56,Alvarez04,DU14}, have also achieved great successes, and high (even exponential) precisions can be obtained. However, comparing  the uniform asymptotic approximation method with them directly  is not an easy task. In particular,  the precisions of these methods, including the asymptotic approximation method,  depend on the models and properties of the potentials $V(x)$ appearing in (\ref{sch}), and for a given model, one method might achieve much better precisions than others, but in other models the situation can be completely different. In the uniform WKB approximations and the current method,  the analysis and precisions of a given problem also depend on the specific choice of the comparison equations \cite{RY64,Slavyanov96,FF96}. Considering the scope of this paper, we shall leave this issue to another occasion. }

 {The new analytical approximation method presented in this paper is also useful to future investigations on} the high-order approximations, extension to the nonlinear Schr\"{o}dinger equation, and applications to various quantum systems with general potentials. Our formulas are general and simple to use, and can also be applied  to other interesting cases,  for example,  Hawking radiation and  quasi-normal modes of black holes \cite{konoplya_quasinormal_2011, berti_quasinormal_2009}, primordial perturbations during inflation and reheating \cite{bassett_inflation_2006}, and Schwinger vacuum pair productions due to laser pulses \cite{dumlu_stokes_2010}.

%%%%%%%%%%%%%%%%%%%%%%%%%%%%%%%%%%%%%%%%%%%
%\vspace{6pt}
%
%%%%%%%%%%%%%%%%%%%%%%%%%%%%%%%%%%%%%%%%%%%
%%% optional
%%\supplementary{The following are available online at \linksupplementary{s1}, Figure S1: title, Table S1: title, Video S1: title.}
%
%% Only for the journal Methods and Protocols:
%% If you wish to submit a video article, please do so with any other supplementary material.
%% \supplementary{The following are available at \linksupplementary{s1}, Figure S1: title, Table S1: title, Video S1: title. A supporting video article is available at doi: link.}
%
%%%%%%%%%%%%%%%%%%%%%%%%%%%%%%%%%%%%%%%%%%%
%\authorcontributions{All authors have equally contributed, read and agreed to the published version of the manuscript.}
%
%%%%%%%%%%%%%%%%%%%%%%%%%%%%%%%%%%%%%%%%%%%
%%\funding{Please add: ``This research received no external funding'' or ``This research was funded by NAME OF FUNDER grant number XXX.'' and  and ``The APC was funded by XXX''. Check carefully that the details given are accurate and use the standard spelling of funding agency names at \url{https://search.crossref.org/funding}, any errors may affect your future funding.}

\section*{Acknowledgements}
This work is partially supported by the National Natural Science Foundation of China under the Grants Nos. 11675143, 11675145, and 11975203;
 the Zhejiang Provincial Natural Science Foundation of China under Grant No. LY20A050002; and the Fundamental Research Funds for the Provincial Universities of Zhejiang, China under Grants No. RF-A2019015.
 B.-F. L is partially supported by NSF grant PHY-1454832.

\end{document}